\documentclass{article}

\usepackage{graphicx}    
\usepackage{cite}     
\usepackage{amsmath,amssymb,amsfonts}
\usepackage{authblk}
\usepackage{times} 
\usepackage{amsmath} 
\usepackage{amssymb}  
\usepackage{dsfont}
\usepackage{subcaption}  
\usepackage{calrsfs}
\usepackage{mathrsfs}
\usepackage{eucal}
\usepackage{amssymb}
\usepackage{amsfonts}
\usepackage{mathtools}
\usepackage{algorithm}
\usepackage{algpseudocode,color}
\usepackage{cite}
\usepackage{physics}
\date{}

\title{A control system framework for counterfactuals: an optimization based approach \footnote{This manuscript is the joint version of two separate works by the authors \cite{ecc2025,Rocond2025cf} }}

\author{
    Pierluigi Francesco De Paola$^{*,1,2,3}$, Jared Miller$^{4}$,
    Alessandro Borri$^{3}$,\\ 
    Alessia Paglialonga$^{2}$, Fabrizio Dabbene$^{2}$
}

\affil{
\small 
    $^1$Politecnico di Bari, Bari, Italy\\
    $^2$Consiglio Nazionale delle Ricerche, Istituto di Elettronica e di Ingegneria dell’Informazione e delle
Telecomunicazioni (CNR-IEIIT), Turin, Italy\\
    $^3$Consiglio Nazionale delle Ricerche, Istituto di  Analisi dei Sistemi ed Informatica (CNR-IASI), Rome, Italy\\
    $^4${Department of Information Technology and Electrical Engineering, ETH, Zurich, Switzerland}\\
    $^*$Corresponding author: pierluigi.depaola@ieiit.cnr.it
}

\begin{document}

\maketitle

\begin{abstract}
Counterfactuals are a concept inherited from the field of logic and in general attain to the existence of causal relations between sentences or events. In particular, this concept has been introduced also in the context of interpretability in artificial intelligence, where counterfactuals refer to the minimum change to the feature values that changes the prediction of a classification model.
The artificial intelligence framework of counterfactuals is mostly focused on machine learning approaches, typically neglecting the physics of the variables that determine a change in class. 
However, a theoretical formulation of counterfactuals in a control system framework - i.e., able to account for the mechanisms underlying a change in class - is lacking. To fill this gap, in this work we propose an original control system, physics-informed, theoretical foundation for counterfactuals, by means of the formulation of an optimal control  problem. We apply the proposed methodology to a general glucose-insulin regulation model and results appear promising and pave the way to the possible integration with artificial intelligence techniques, with the aim of feeding machine learning models with the physics knowledge acquired through the system framework. 
\end{abstract}

\section{Introduction}
In the context of artificial intelligence (AI) classification problems, counterfactuals can be defined as the minimum  change that should occur in an instance to observe a different outcome from the classifier \cite{guidotti2024counterfactual}.
From a theoretical perspective, counterfactuals suggest what should have been different in an instance (defined as \textit{factual}) to vary with minimum change or \textit{minimum effort} \cite{guidotti2024counterfactual} the output class of the AI model. In this sense, the concept of counterfactual encodes the idea of 'what-if', clearly expressed by Lewis in the sentence "if Kangaroos had no tails, they would topple over" \cite{lewis2013counterfactuals}.
Counterfactuals have been widely applied to classification problems in various applications, for example in health prediction and prevention where different classes are typically associated with the risk of developing a given disease, or in safety-critical applications, with classes representing safe and unsafe sets \cite{wachter2017counterfactual,carlevaro2022counterfactual,lenatti2022novel,carlevaro2023multi}.\\
Nevertheless, although providing explainable insights into the decision of the algorithm, the \textit{AI-driven} counterfactuals are typically based on classifiers trained on data (i.e., input-output associations) and may neglect the underlying dynamics of the system, consequently providing merely conceptual decisions \cite{guidotti2024counterfactual,stepin2021survey,kment2006counterfactuals}. The aim of this study is to introduce a control system theoretical formulation for counterfactuals, with the aim of assessing a physics-informed approach suitable to account for the underlying mechanisms driving the change of class. 
The concept of counterfactual in the control systems framework is introduced by means of an optimal control problem, aimed at computing the minimum control law steering a given initial condition from a given set of the state space (e.g. the unsafe set) to another one (e.g. the safe set).
Moreover, after introducing the concept for the case of perfectly known systems, we discuss in this work also how the proposed approach 
can be extended to account for uncertainties on the knowledge of the parameters of the system under consideration.
This work carries on a line of research by the authors aimed at leveraging advanced learning and control methods for life science and biological applications. Biological systems are highly nonlinear and prone to uncertainties that can turn out to be critical in many real-world applications. Hence, defining a general ``robust" framework 
for control-driven counterfactuals is crucial for the reliability of the proposed approach.
This problem is then here cast as an infinite-dimensional problem in the space of measures and subsequently solved by means of the moment-sum-of-squares (moment-SOS) hierarchy through a sequence of convex relations in the space of the moments \cite{henrion2008nonlinear,lasserre2008nonlinear,henrion2020moment,lasserre2009moments}, 
with the aim of deriving a general methodology suitable to be exploited for both linear and nonlinear systems.
This work is preliminary to the integration of control and AI methods to derive physics-informed personalized minimum recommendations for disease prevention.

\section{METHODOLOGY}

\subsection{Mathematical background}
\(\mathbb{R}^n\) denotes the n-dimensional real euclidean space and  \(\mathbb{R}[x]\) the ring of polynomials in the variable $x$. \(C({X})\) is the vector space of continuous functions and \( \mathcal{M}({X}) \) constitutes its dual space of Borel measures on $X$, with $X$ subset of \(\mathbb{R}^n\). $C(X)$ and \(\mathcal{M}({X}) \) are paired via the continuous linear functional $Tf=\int_X f(x) d\mu(x)$.
$\mathcal{M_{+}}(X) $ and $C_{+}(X)$ denotes respectively the nonnegative cones of $\mathcal M(X)$ and $C(X)$. $C^1(X) \subset C(X)$ is the set of continuous functions with continuous first derivatives. A set $X \subseteq \mathbb{R}^n$ of the form $ X = \{x\mid s_k(x) \geq 0\}$ is said to be semialgebraic. A polynomial $\pi(x)$ can be expressed in the monomial basis as $\sum_\alpha \pi_\alpha x^\alpha$ and the degree of $\pi(x)$ ($\text{deg}(\pi(x)$)) is considered to be the maximum $|\alpha|= \sum_i |\alpha_i|$.

The indicator function of  set $A \subseteq X$ is a function $\mathds{I}_A :X \to \{0,1\}$ such that $\mathds{I}_A(x)=1$ if $x \in A$ and $\mathds{I}_A(x)=0$ otherwise and the measure of a set $A$ with respect to $ \mu \in \mathcal{M}(X)$ is defined as $\mu(A) = \langle \mathds{I}_A(x), \mu \rangle$.

The {$\alpha$}-order moment of a measure $\mu$ is the scalar quantity $y_\alpha= \langle x^\alpha, \mu \rangle$. For $\alpha=1$ we get the mass of the measure, i.e. $\mu$ is $\langle 1, \mu \rangle$.\\

A Dirac delta measure or \textit{atomic} measure \(\delta_{x=x'}\in \mathcal{M}(X)\) (also referred to as \(\delta_{x'}\) in the following) is a measure supported on a single point (i.e.,\textit{atom}) $x=x'$.
We denote the support of a measure $\mu$ as spt$(\mu)$.
The product $\mu_{1}\otimes\mu_{2}$ denotes the product of the measures $\mu_{1}$ and $\mu_{2}$ that satisfies $(\mu_1 \otimes \mu_2)(A \times B)=\mu_1(A)\mu_2(B)$. \\ 

A given linear operator ${H} : X \to X$ is endowed with  with a unique adjoint operator ${H}^*$ such that ${H}^*=H$, that is $\langle Hv, \mu \rangle = \langle v, H^*\mu \rangle$, $\forall \ {v}(t,x)\in C^1([0,T] \times X) \in X$, $ \forall \ \mu \in \mathcal{M}(X)$. $\mathcal{L}_f$ will denote throughout the text the Lie derivative of a general function with respect to the vector field $f$.

\subsection{Occupation Measures}
Occupation measures are great tools for solving nonlinear optimization problem for dynamical systems\cite{MillerEtAl},\cite{case2024}.
Let assume that \(X \subseteq \mathbb{R}^n\) is the state space of a dynamical system, and let \( x(t \mid x_0) \)  denote the trajectory of the system starting from initial condition \(x_0\) over the time interval. The occupation measure $\mu$ of a set $A \times B$ with $A \subseteq [0,T] $ and $B \subseteq X $ referred to the trajectory $x(t \mid x_0)$ is defined as 
\[
    \mu(A \times B \mid  x_0) \coloneqq \int_{0}^{T} \mathds{I}_{A\times B}(t, x(t \mid x_0)) \, dt
\]
and can be interpreted as the time spent by the trajectory in the set $A \times B$. \\
Given a distribution of initial conditions $\mu_0 \in  \mathcal{M}(X_0)$ the occupation measure $\mu$ is averaged over the initial distribution $\mu_0$ and yields the $\mu_0$-averaged amount of time that the trajectories dwell in $A \times B$

\[
    \mu(A \times B) \coloneqq \int_{X_0} \mu(A \times B \mid  x_0)  \, d\mu_0(x_0).
\]

Analogously, the final measure $\mu_T$ represents the distribution of the final states as they are transferred by the dynamics according to the initial distribution $\mu_0$,
\[
    \mu_T(B) \coloneqq \int_{X_0} \mathds{I}_{B}(x(T|x_0))d\mu_0{(x_0)}.
\]

The three measures $\mu_0, \mu, \mu_T$ are linked together by the Liouville's equation  

\[
\langle v(T,x), \mu_T \rangle = \langle v(0,x), \mu_{{0}} \rangle + \langle \mathcal{L}_fv(t,x), \mu \rangle
\]

that admits also a short-hand formulation independent from the general test function ${v} \in C^1([0,T]{\times X})$ by means of the linear adjoint operator $\mathcal{L}^*$,

\[
 \delta_T \otimes \mu_T= \delta_0 \otimes \mu_0 + \mathcal{L}^* \mu.
\]

Liouville's equation can be interpreted to encode the system dynamics in the measure formulation.

Every trajectory $x(t|x_0)$ with $x_0 \in  X_0$ and $t \in [0,T]$ induces a measure formulation with $\mu_0 = \delta_{{x_0}}$ and $\mu_T = \delta_{{x_{T \mid x_0}}}$.

\subsection{Moment-SOS hierarchy}
A standard formulation for an infinite-dimensional LP with measure variables $\mu \in \mathcal M_+(X)$ is the following \cite{lasserre2009moments}

\begin{equation}
\begin{aligned}
p^* = & \underset{\mu \in \mathcal{M}_+(X)}{\text{inf}} 
& & \langle p(x), \mu \rangle \\
& \quad \text{s.t.}
& & \langle a_{i}(x), \mu \rangle = b_i, \quad \forall i = 1, \ldots, I_{\max}
\end{aligned}
\label{eq:gen_infinite_dim}
\end{equation}

Here $p(x) \in C(X)$ is the cost functional and $\langle a_{i}(x), \mu \rangle = b_i$ are a set of  affine constraints, with  $a_i \in  C(X)$.
A dual program can be formulated over the space of nonnegative continuous functions in the form \cite{nash1987linear}:
\begin{equation}
\begin{aligned}
q^* =& \underset{v \in \mathbb R^{m}}{\text{sup}} 
& & \sum_i b_iv_i\\
& \quad \text{s.t.}
& & p(x) -\sum_i a_i(x)v_i \geq 0, \\  & & & \forall i = 1, \ldots, I_{\max}, x\in X
\end{aligned}
\label{eq:general_dual}
\end{equation}

Under the assumption of $p(x),a_i(x)$ polynomials and $X$ semialgebraic, the program (\ref{eq:gen_infinite_dim}) can be  discretized for tractable computation and restated as an infinite dimensional LP in the space of the moments of the measure $y_{\alpha}$ and in turn truncated in finite dimensional relaxations in the space of moments up to a given relaxation order  \textit{d}, such that $2d \geq \text{max} (\text{deg}(f(x),\text{deg}(g_i(x))$ as follows

\begin{subequations}
\begin{align}
p^*_{d} = \ & \underset{y}{\text{min}} 
& & \sum_{\alpha} p_{\alpha}y_{\alpha} \\
& \quad \text{s.t.}
& & \sum_{\alpha} a_{i\alpha}y_{\alpha}= b_{i}, \quad \forall i = 1, \ldots, I_{\max} \\
& & & \mathbb{M}_d(y) \succeq 0, \label{eq:2c} \\
& & & \mathbb{M}_{d-d_k}(s_ky) \succeq 0, \label{eq:2d}\quad \forall k = 1, \ldots, N
\end{align}
\label{eq:relaxed}
\end{subequations}

$\mathbb{M}_d(y),\ \mathbb{M}_{d-d_k}(w_ky)$ (Moment and the Localizing Matrix) enforce  constraints (\ref{eq:2c})-(\ref{eq:2d}) in the space of truncated moments and guarantee that $y$ are sequences of moments representing measure.
The moment-SOS hierarchy for $d$-degree solutions to problem yields sequence of increasing lower bounds to the optimum $p^*$, $p^*_{d} \leq p^*_{d+1} \leq p^*_{d+2} \cdots$, with convergence guaranteed as $d \to \infty $ under suitable compactness properties of the support set $X$.\\

\textit{Remark}. The Moment-SOS hierarchy is a valuable method to approximate the solution of infinite dimensional program in measures. Other possible methods are discussed in \cite{fattorini1999infinite}.

\section{Fully parametrized systems}
\subsection{Problem statement}
Consider a general optimal control problem with free terminal time  in the form:

\begin{subequations}
\begin{align}
    P^*=\min_{{u},\ \tau{\in[0,T]}} \quad & J({x,u}) = \Phi(x(\tau)) + \int_0^\tau \phi(x(t), u(t)) \, dt \label{eq:objective_known} \\
    \text{s.t.} \quad & \dot{x}(t) = f(x(t), u(t)),\quad {t \in [0,{\tau}]},  \label{eq:dynamics_known} \\
    & x(0) \in X_0, \  x(t) \in X, \quad {t \in [0,{\tau}]},\label{eq:state_constraints_known} \\
    & x(\tau) \in X_T,\label{eq:terminal_condition_known} \\
    & u(t) \in U, \label{eq:control_constraints_known}\quad t \in [0,{\tau}],
\end{align}
\end{subequations}
where \(x(t) \in \mathbb{R}^n\) represents the state of the system at time \(t\), \(u(t) \in \mathbb{R}^m\) represents the control input, and \(f: \mathbb{R}^n \times \mathbb{R}^m \to \mathbb{R}^n\) is the vector field describing the system dynamics, \(\tau\) is the free terminal time, \(\Phi(x(\tau))\) is the terminal cost, a function of the final state \(x(\tau)\), \(\phi(x(t), u(t))\) is the running cost, a function of the state \(x(t)\) and of the control \(u(t)\). $X_0,\ X, \ X_T$ represent respectively the set of initial conditions, the set of the trajectories and the set of terminal states, $T$ is the time horizon. \\
Within the framework of the optimal control problem outlined above, in this study we define the concepts of factual and counterfactual (CS-factual and CS-counterfactual) as follows.\\
\textit{Definition 1} (\textit{CS-factual}). 
Consider a general optimal control problem of the form (\ref{eq:objective_known})-(\ref{eq:control_constraints_known}). We define the CS-factual as 
\[
  x_{\text{f}} \coloneqq x(0) \in X_0.
\]
\textit{Definition 2} (\textit{CS-counterfactual}).
Consider a general optimal control problem (\ref{eq:objective_known})-(\ref{eq:control_constraints_known}) and assume that a solution $u^*(\cdot) = \arg\min_{u(\cdot), \, \tau} J(x(t),u(t))$ exists.
Given a CS-factual $x_{\text{f}} \in X_0$, we define the CS-counterfactual $x_{\text{cf}}$ associated with the factual $x_{\text{f}} \in X_0$ as
\[
  x_{\text{cf}} \coloneqq x(\tau \mid x_{\text{f}},{u^*}), \ x_{\text{cf}} \in X_T,
\]
\\
{where $x(t \mid x_0,u)$ denotes the state reached at time $t$ from $x_0$ with input function $u$.} \\
It can be observed that, by definition, the \textit{CS-counterfactual} encodes the concept of "\textit{minimum effort}" for a vector field to access the terminal set $X_T$.
For the purpose of this study we will consider the following assumption.\\
\textit{Assumption 1}. The set $X_0 \subseteq X_u$ and the set $X_T \subseteq X_s $ where $X_u$ and $X_s$ are respectively the unsafe and the safe set for the system. \\
From now on the the terms "factual" and "counterfactual", when clear from the context, will specifically refer to the CS-factual and CS-counterfactual.
The following part of the work illustrates how it is possible to extract counterfactuals leveraging occupation measures and moment-SOS hierarchy.

\subsection{Problem solution in measure space}
Optimal control problems of the form (\ref{eq:objective_known})-(\ref{eq:control_constraints_known}) can be bounded by infinite dimensional LP in measures as the following \cite{henrion2008nonlinear,lasserre2008nonlinear,henrion2020moment}
\begin{subequations}
\begin{align}
    p^* =\inf_{\mu,\mu_\tau} \quad & \langle \phi, \mu \rangle + \langle \Phi, \mu_\tau \rangle  \label{eq:functional_known} \\
    \text{s.t.} \quad  & \mathcal{L}^{*} \mu = \delta_0 \otimes \mu_0 - \mu_\tau \label{eq:Liouville_known}  \\
    & \mu \in \mathcal{M_+}([0,\tau] \times X \times U), \label{eq:support1_known} \\
    &\mu_\tau \in  \mathcal{M_+}(X_T),\label{eq:support2_known} 
\end{align}
\end{subequations}
where $\mu_0 = \delta_{{x_0}}$ is the initial measure with support on the initial condition $x_0$, $\mu$ is the occupation measure of the system trajectories, whereas $ \mu_\tau = \delta_{\tau} \otimes \delta_{{x_\tau}}$ generalizes the concept of terminal measure with free terminal time. It holds that $\langle 1, \mu \rangle = \tau$.
The dual problem in the space of the continuous function reads as: 
\begin{subequations}
\begin{align}
    q^* = \sup_{\upsilon \in \mathbb{R}} \quad & \upsilon
    \label{eq:functional_dual_known} \\
    \text{s.t.} \quad  & \upsilon \geq v(x), \ \forall x\in X_0 \\
    & \mathcal{L}_fv(x) \leq -\phi(x),\ \forall (x,u) \in X \times U \notag\\
    & v(x) \leq \Phi(x),\ \forall x \in X_T \label{eq:support_dual_known}  \\
    & v \in C^1(X).\label{v_function_known}
\end{align}
\label{eq:dual_known}
\end{subequations}
The polynomial $v(x(t))$ that solves (\ref{eq:functional_dual_known})-(\ref{eq:support_dual_known}) provides a polynomial subsolution of the Hamilton-Jacobi-Bellman equation which approximates the value function along all the optimal trajectories of the system. \\
The following assumptions are made in program (\ref{eq:objective_known})-(\ref{eq:control_constraints_known}) for the development of this work:\\
\textit{Assumption 2}. The vector field $\dot{x}= f(x(t),u(t))$ is Lipschitz in each argument in the compact set $[0,T] \times X \times U$  and is polynomial, thus $f(x(t),u(t)) \in \mathbb{R}^n[x,u]$. $X, \ X_T$ are semialgebraic sets, $X= \{ x \mid w_k(x) \geq 0, {k= 1,\cdots, N}\}$, $X_T= \{ x \mid w_{Tk}(x) \geq 0, {k= 1,\cdots, N}\}$, $deg(w_k(x))=deg(w_{Tk}(x))=d_k$ {for all $k$}; \\
\textit{Assumption 3}. The vector field is affine in the control input, $\dot{x}= h(x(t))+ g(x)u$; \\
\textit{Assumption 4}. The class of admissible controls $u$ is $L^p$-measurable, i.e.
\[
\|u(t)\|_{L^p,[0,T]} = \left( \int_{0}^T \|u(t)\|^p \, dt \right) < \infty, \quad p \in [1,\infty).
\]
\textit{Assumption 5}. The \textit{effort} of the vector field is considered to be the ${L^2}$-norm of the control input $u$, $\|u(t)\|_{{L^2([0,T]})} = \left( \int_{0}^{T} \norm{u}^2 \, dt \right)$; no terminal cost is considered in the functional (\ref{eq:objective_known}). 
\\
As a consequence, for the development of the proposed methodology, the general formulation of ({\ref{eq:functional_known})-(\ref{eq:support2_known}}) takes the form:

\begin{subequations}
\begin{align}
    p^* =\inf_{\mu,\mu_\tau} \quad & \langle u^2, \mu \rangle   \label{eq:functional_ac_known} \\
    \text{s.t.} \quad  & \mathcal{L}^{*} \mu = \delta_0 \otimes \mu_0 - \mu_\tau \label{eq:Liouville_ac_known}  \\
    & \mu \in \mathcal{M_+}([0,\tau] \times X \times U), \label{eq:support1_ac_known} \\
    &\mu_\tau \in  \mathcal{M_+}(X_T). \label{eq:support2_Ac_known} 
\end{align}
\label{eq:AC_known}
\end{subequations}

\subsection{LMI formulation}  
The problem in (\ref{eq:functional_ac_known})-(\ref{eq:support2_Ac_known}) is solved in the space of moments with relaxation order $d$ with the following LMI formulation \cite{lasserre2008nonlinear,henrion2020moment} :

\begin{subequations}
\begin{align}
p^*_{d} = \ & \underset{y,y_\tau}{\text{min}} 
& & y_{02} \label{eq:cost_LMI_known} \\
& \quad \text{s.t.}
& & Liou_\alpha(y_0,y,y_\tau)\ \forall \alpha = 1, \ldots, 2d \label{eq:Liou_LMI_known}\\
& & & \mathbb{M}_d(y) \succeq 0, \ \mathbb{M}_d(y_\tau) \succeq 0\label{eq:2c_LMi_known} \\
& & & \mathbb{M}_{d-d_k}(w_ky) \succeq 0, \quad \forall k = 1 \ldots, N \\
& & & \mathbb{M}_{d-d_k}(w_{Tk}y_\tau) \succeq 0,   \quad \forall k = 1 \ldots, N_{T} \label{eq:2d_LMI_known}, 
\end{align}
\label{eq:LMI_known}
\end{subequations}

where $y_0,y,y_\tau$ represent the moment sequence respectively of measures $\mu_0, \mu, \mu_\tau$, $y_{02}$ is the cost in (\ref{eq:functional_ac_known}) expressed as $2^{nd}$- order moment of the control input $u$ with respect to the occupation measure $\mu$. Constraint (\ref{eq:Liou_LMI_known}) represents the Liouville equation in (\ref{eq:Liouville_ac_known}) in the moment space. Moment and Localizing matrices in constraints (\ref{eq:2c_LMi_known})-(\ref{eq:2d_LMI_known}) require moments up to degree $2d$ and enforce the measure support contraints in (\ref{eq:support1_ac_known})-(\ref{eq:support2_Ac_known}).

\subsection{Counterfactual Extraction}
In this section we illustrate two algorithms that can be exploited independently to extract counterfactuals by means of the methodology illustrated above. \\
\textit{Proposition 1}. Consider the problem 
in (\ref{eq:functional_ac_known})-(\ref{eq:support2_Ac_known}) and its $d$-degree solution  and consider a factual $x_\text{f}=\text{spt}(\mu_0=\delta_{{x_\text{f}}})$. If $\text{mass}(\mu_\tau)=1$, then  the counterfactual $x_\text{cf}=\text{spt}(\mu_\tau=\delta_{{x_\tau}})$ and it holds that
$x_\text{cf}=y_{1,\tau}$, where $y_{1,\tau}$ denotes the $1^{st}$-order moment of the final measure $\mu_{\tau}$. \\ 

Hence, counterfactual $x_\text{cf}$ can be extracted from the Moment Matrix of the final measure $\mathbb{M}_d(y_\tau)$. Algorithm 1 summarizes this procedure. \\
\textit{Remark}. If (\ref{eq:AC_known}) admits $r$ optimal solutions, then the final measure $\mu_{\tau}$ is r-atomic (i.e. $\mu_{\tau}= 
\sum_{i=1}^{r} \beta_i \cdot \delta_{x_{\tau_i}}$). Hence, the counterfactual $x_\text{cf}=\text{spt}(\mu_\tau=\delta_{x_\tau})= \sum_{i=1}^{r} \beta_i \cdot {x_{\tau_i}}$.  \\
Moreover, considering the dual formulation (\ref{eq:dual_known}), the following alternative method holds. \\
\textit{Proposition 2}. Consider the problem 
in (\ref{eq:functional_ac_known})-(\ref{eq:support2_Ac_known})  and its dual formulation in the form (\ref{eq:dual_known}) with a given $d$-degree solution that yields a subsolution $v(x)$ of the value function. Let $u^*(x) := \arg\min_{u \in U} \left\{ \mathcal{L}_f v(x) + \phi(x, u) \right\}$ and let $\tau=\text{mass}(\mu)$. Then, for a given $x_{\text{f}} \in X_0$, 
$x_{\text{cf}}=x(\tau|x_{\text{f}},u^*(x))$. \\

Algorithm 2 summarizes the procedure. 
For input-affine control system and running cost of the form $\phi(x,u)= \phi_x(x) +u^Tu$, the control law $u^*(x)$ can be derived from the first order optimality condition \cite{lasserre2008nonlinear}\cite{henrion2020moment}:
\begin{align}
&\frac{\partial (\phi(x,u) + \nabla v(x) \cdot (h(x)+g(x)u))}{\partial u} = 2u + \nabla v(x)u = 0 \\ 
&\Rightarrow  u^*(x)= -\frac{1}{2} \nabla v^*(x)\cdot g(x). \label{u_star}
\end{align}
Proposition 1 and Proposition 2 follow almost directly from the developments in \cite{henrion2008nonlinear,lasserre2008nonlinear}.\\
\begin{algorithm}
\caption{Counterfactual extraction via Moment Matrix}
\begin{algorithmic}[1]
\Require Sets \(X_0\), \(X\), \(X_T\), dynamics \(f\), cost functional, initial degree \(d\)
\Ensure Factual-Counterfactual pair generation \\ 
Define $\mu_0$, (factual $x_\text{f}=\text{spt}(\mu_0=\delta_{{x_\text{f}}})$)
  \State Solve (\ref{eq:AC_known}) at degree \(d\)
  \State Extract $y_{1,\tau}$ from $\mathbb{M}_d(y_\tau)$
  \State $x_\text{cf}=y_{1,\tau}$   
\end{algorithmic}
\end{algorithm}

\begin{algorithm}
\caption{Counterfactual extraction via trajectory recovery}
\begin{algorithmic}[1]
\Require Sets \(X_0\), \(X\), \(X_T\), dynamics \(f\), cost functional, initial degree \(d\)
\Ensure Factual-Counterfactual pair generation \\ 
Define $\mu_0$, factual $x_\text{f}=\text{spt}(\mu_0=\delta_{{x_\text{f}}})$
 \State Solve (\ref{eq:AC_known}) at degree \(d\)
\State Extract $\tau=\text{mass}(\mu)$
\State Retrieve $v(x)$ from the dual program (\ref{eq:dual_known}) in the SOS version
 \State Retrieve $u^*(x) := \arg\min_{u \in U} \left\{ \mathcal{L}_f v(x) + \phi(x, u) \right\}$
 \State Simulate the system in closed-loop $\dot{x}=f(x(t),u^*(x(t))$
\State $x_{\text{cf}}=x(\tau|x_{\text{f}},u^*(x))$
\end{algorithmic}
\end{algorithm}

{\textit{Remark}}. 
As discussed in Section II, solving (\ref{eq:functional_known})-(\ref{eq:support2_known}) and (\ref{eq:functional_dual_known})-(\ref{v_function_known}) for a given relaxation order $d$ yields  lower bound  $p^*_{d}=q^*_{d} \leq p^*=q^*$ and the convergence is guaranteed as $d \to \infty$.
Moreover, as $d \to \infty$, the following holds. \\
\textbf{Theorem 1}. Consider the problem (\ref{eq:functional_ac_known})-(\ref{eq:support2_Ac_known}) with dual in the form (\ref{eq:functional_dual_known})-(\ref{v_function_known}) and its solution for a given relaxation order $d$. As $d \to \infty$, the trajectory $(x(t)|x_\text{f},u_d^*(x))$ leading to the counterfactual $x_\text{cf}$ is bounded.\\
\textit{Proof}. From Theorem 4.1 \cite{lasserre2008nonlinear},  as $d \to \infty$, $v_d(x) \to v_{\inf}(x):= \mathcal{V}(x)$ value function. \\
Consequently, $ 0\leq \mathcal{V}(x) \leq \mathcal{V}(x(0))$. Moreover, from (\ref{eq:functional_dual_known})-(\ref{v_function_known}), the sets 
\[
 \Omega_0=\{{\mathcal{V}(x)} \geq 0 \} \ \text{and} \ \Omega_{V_0}=\{{\mathcal{V}(x)} \leq \mathcal{V}(x(0))\} 
\]

are positively invariant. 
Hence, there exists a function \cite{khalil2002nonlinear} 
$\beta \in \mathcal{KL}$ such that 
\[
\|x(t)\| \leq \beta(\|x(t_0)\|,t-t_0) = \beta(\|x_\text{f}\|,t-t_0).
\]
Moreover, we will prove the following:\\
\textbf{Theorem 2}. Consider the problem (\ref{eq:functional_ac_known})-(\ref{eq:support2_Ac_known}) with dual in the form (\ref{eq:functional_dual_known})-(\ref{v_function_known}) and its solution for a given relaxation order $d$. As $d \to \infty$, the terminal set $X_T$ containing the counterfactual $x_\text{cf}$ is asymptotically stable.\\
\textit{Proof}. By assumptions in program (\ref{eq:functional_dual_known})-(\ref{v_function_known})
$v(x) \in C^1(X)$
Moreover from Theorem 4.1 \cite{lasserre2008nonlinear},  as $d \to \infty$, $v_d(x) \to v_{\inf}(x):= \mathcal{V}(x)$ value function and by assumptions  in program (\ref{eq:functional_dual_known})-(\ref{v_function_known})
$\mathcal{V}(x)=0$ in $X_T$ and $\mathcal{V}(x)>0$ elsewhere. Since by (5c) it holds that
\[
\mathcal{L}_f\mathcal{V}(x) \leq -\phi(x(t)),
\]
with $\phi(x)= u^2(x)$, being $u^2(x) \in \mathcal{K}_\infty$  by Lyapunov direct method the asymptotic stability follows.

\textbf{Theorem 3}: Consider the problem (\ref{eq:functional_ac_known})--(\ref{eq:support2_Ac_known}) and its dual (\ref{eq:functional_dual_known})--(\ref{eq:support_dual_known}) in the SOS form. Let assume it is feasible  and that a solution exists for all suitable relaxation orders $d$. Let denote with $\eta :S \times S \to \mathbb R$ a general metric on the euclidean space $\mathbb{R}^{n}$ ($S \subseteq \mathbb{R}^{n})$. As $d \to \infty$, the distance $\eta({x_{\text{cf},d}},{x_{\text{cf},\infty}})$ is bounded. \\
\textit{Proof.} By assumption $X_T$ is a compact set in $\mathbb {R}^{n}$. 
Being $X_T$ compact, $X_T \times X_T$ is also compact, being the product of compact spaces compact in the standard topology. Moreover, the function $\eta$ is  continuous on $X_T \times X_T$ by definition.
From the extreme value theorem, a continuous (real-value) function on a compact set attains a maximum and minimum, which implies the boundedness.
\subsection{RESULTS}
The proposed methodology for counterfactual extraction via occupation measures is applied to a general glucose-insulin regulation  model, the well acknowledged model by Bergman et al. \cite{BergmanEtAl1979}, described by the following system of differential equations:

\begin{subequations}
    \begin{align}
        &\dot{x}_1 = -p_1 x_1 - x_2 x_1 + p_1 G_b \\
        &\dot{x}_2 = -p_2 x_2 + p_3 x_3 \\
        &\dot{x}_3 = -n x_3 + p_4 u 
    \end{align}
    \label{Bergman}
\end{subequations}
where:
\begin{itemize}
    \item ${x_1}$ is the blood glucose concentration, also denoted as $G$ in the following (mg/dl);
    \item ${x_2}$ is the remote insulin concentration ($\mu$U/ml);
    \item ${x_3}$ is the serum insulin concentration, also denoted as $I$ in the following ($\mu$U/ml);
    \item $u$ represents exogenous insulin administration ($\mu$U/ml/min).
\end{itemize}

\begin{figure}[!ht]
\hspace*{-0.5 cm}
\centering
{\includegraphics[scale=0.5]{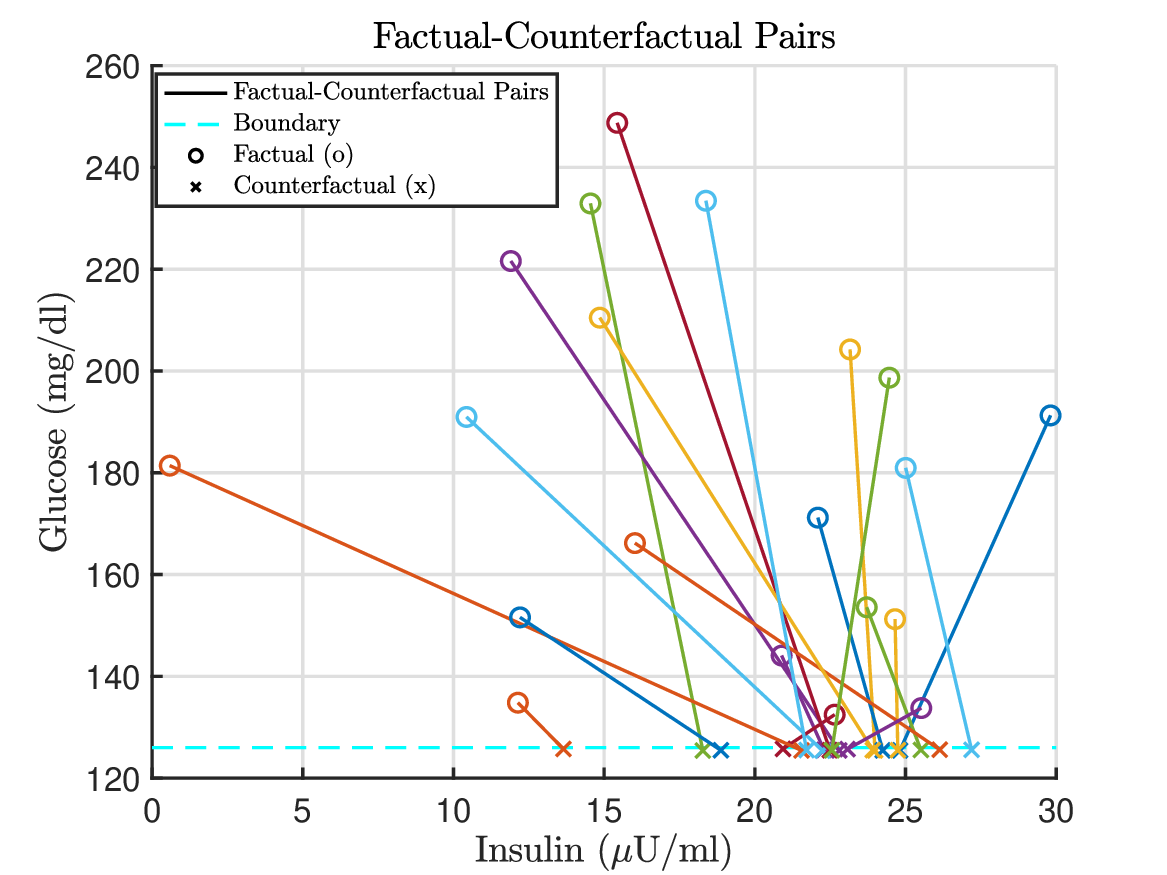}}
\caption{Factual-counterfactual pairs as obtained by solving the optimization problem in the moment space with relaxation order $d=4$.}
\label{fig:coloured_pairs}
\end{figure}

\begin{figure*}[!ht]
{\includegraphics[scale=0.5]{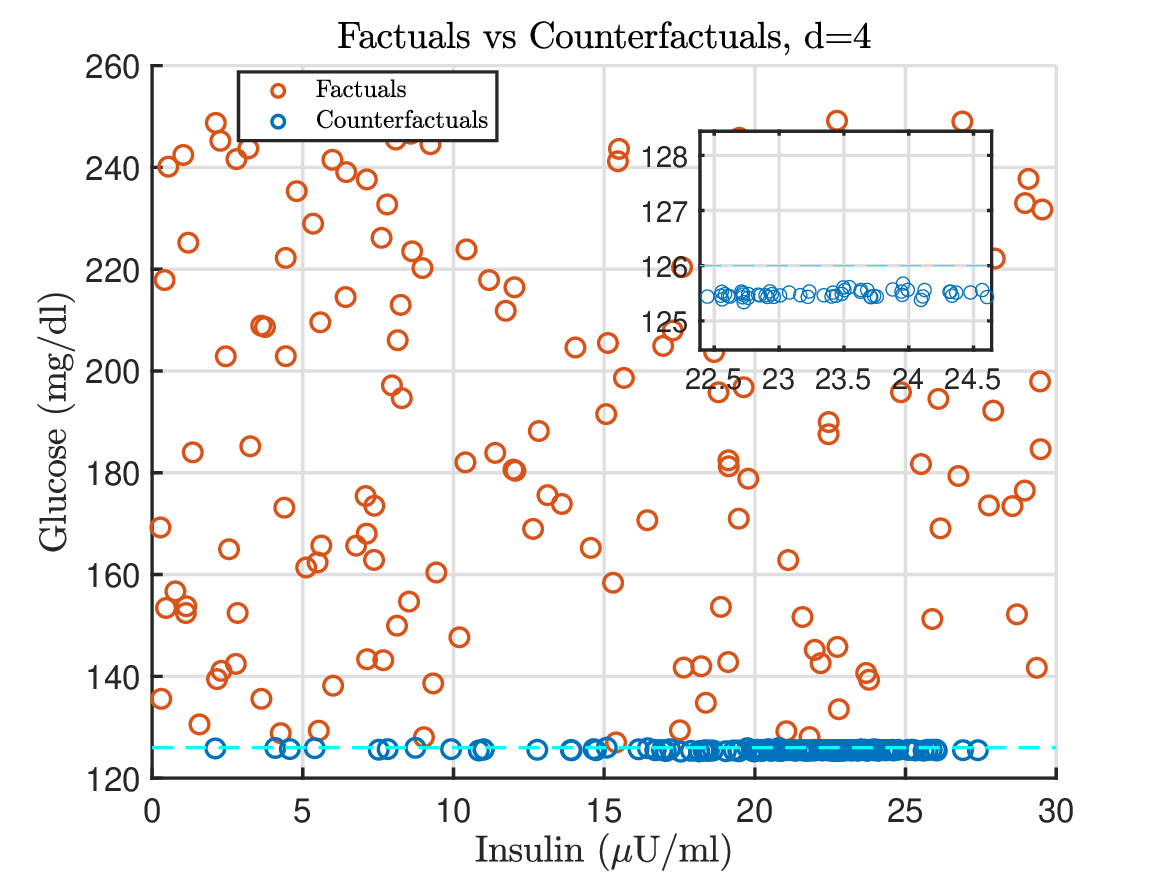}}
{\includegraphics[scale=0.5]{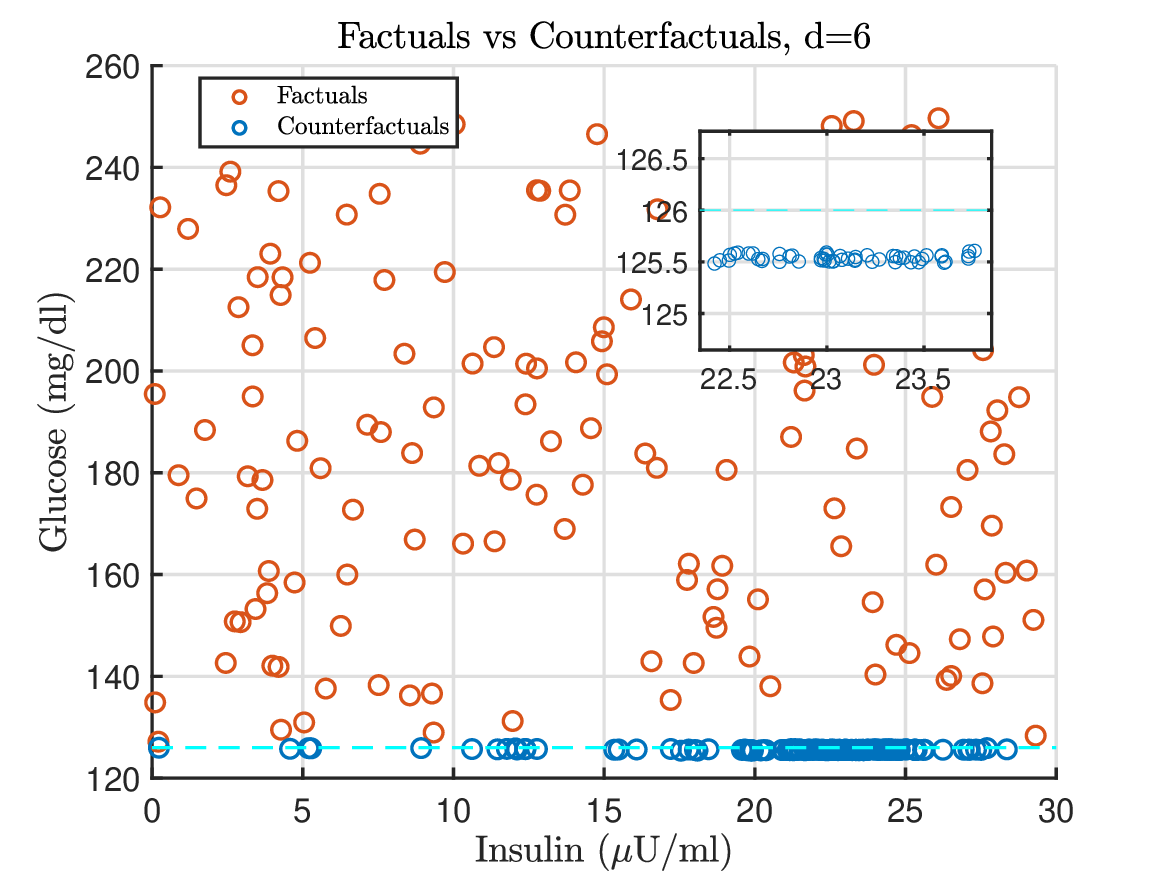}}
\caption{Factuals (red dots) and counterfactuals (blue dots) obtained by solving the optimization problem in the moment space with relaxation order $d=4$ (top panel) and $d=6$ (bottom panel). }
\label{fig:DenseRegions_known}
\end{figure*}

\begin{figure}[!ht]
\hspace*{-0.5 cm}
\centering
{\includegraphics[scale=0.5]{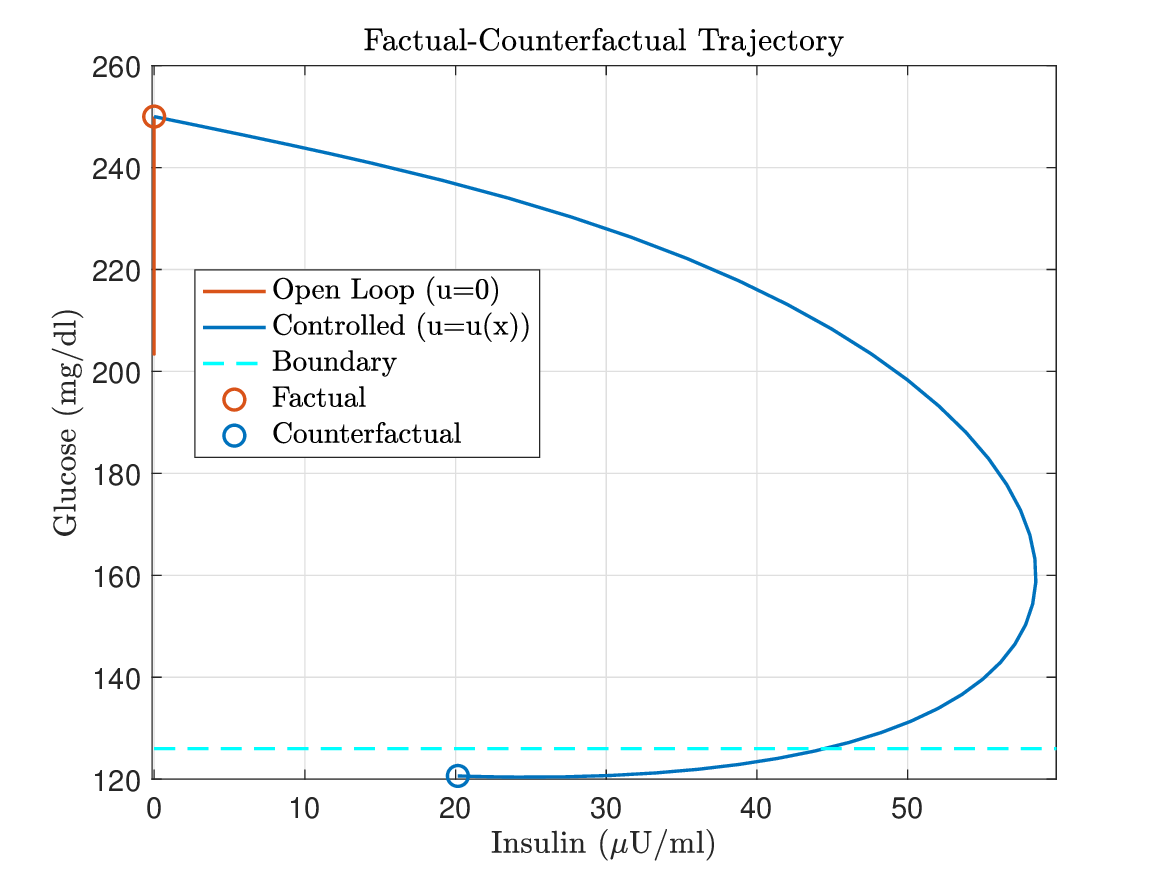}}
\caption{Example of counterfactual trajectory as retrieved for the factual (250,0,0) by applying Algorithm 2.}
\label{fig:trajectory}
\end{figure}

For the values of the parameters reference is made to \cite{BergmanEtAl1979}.
To improve the numerical behavior of the SDP solvers, variables should be normalized, i.e. scaled within boxes [0,1] (see e.g. \cite{henrion2008nonlinear,lasserre2008nonlinear,henrion2020moment}).
Hence, by defining the scaled variables $\hat{x}_1=\frac{x_1}{x_{1_{max}}}$, $\hat{x}_2=\frac{x_2}{x_{2_{max}}}$, $\hat{x}_3=\frac{x_3}{x_{3_{max}}}$ the system results as follows:
\begin{subequations}
    \begin{align}
        &\dot{\hat{x}}_1 = -p_1 \hat{x}_1 - \hat{x}_2 \hat{x}_1 x_{2_{max}} + \frac{p_1 G_b}{x_{1_{max}}} \label{x1hat}\\
        &\dot{\hat{x}}_2 = -p_2 \hat{x}_2 + p_3 \hat{x}_3 \frac{x_{3_{max}}}{x_{2_{max}}}\\
        &\dot{\hat{x}}_3 = -n \hat{x}_3 + \frac{p_4 u}{x_{3_{max}}} \label{x3hat},
    \end{align}
    \label{Bergman_scaled}
\end{subequations}
where $x_{1_{max}}=600$ mg/dl, $x_{2_{max}}=1$ $\mu$U/ml, $x_{3_{max}}=100$ $\mu$U/ml represents reasonable maximum values respectively for the variables $x_{1_{max}}$, $x_{2_{max}}$, $x_{3_{max}}$ \cite{BergmanEtAl1979}. 
 Moreover, time is scaled by multiplying the system (\ref{x1hat})-(\ref{x3hat}) by the time horizon $T$, here assumed to be $T=80\cdot60$ (sec), which is a time range consistent with the model by Bergman et al. \cite{BergmanEtAl1979}. The terminal set is considered to be
$X_T = \{{x} = (x_1, x_2, x_3) \in \mathbb{R}^3 \mid 80 \leq x_1 < 126 \}$\cite{WhelanEtAl2022}, whereas the set of initial conditions is assumed to be $X_0 = \{{x} = (x_1, x_2, x_3) \in \mathbb{R}^3  \mid (126 \leq x_1 \leq 260), (0\leq x_3 \leq 30)\}$\cite{BergmanEtAl1979}. \\
Fig.\ref{fig:coloured_pairs} illustrates factual-counterfactual pairs as obtained when randomly sampling 20 factuals over $X_0$, solving the optimization problem for $d=4$ and extracting counterfactuals via Algorithm 1.
The plot identifies the association between each factual and the related counterfactual in the phase plane ($G$, $I$). Counterfactuals lie in the safe set in the vicinity of the boundary (i.e., the diabetic threshold) separating the safe set from the unsafe set.\\
Fig.\ref{fig:DenseRegions_known} shows the results obtained applying Algorithm 1 when factuals (red dots) are distributed as 150 random initial conditions over the set $X_0$ whereas counterfactuals (blue dots) are extracted solving the optimization problem for two different relaxation orders, $d=4$ (top panel) and $d={6}$ (bottom panel). Both plots show a remarkable feature of counterfactuals, related to the fact that most of the counterfactuals cluster in a dense region within an interval of values of $I$ between 20 and 25 $\mu$U/ml.
This is a notable analogy with respect to the AI-driven framework, that similarly denotes  presence of dense regions of counterfactuals obtained by means of the classification algorithms, as  discussed in \cite{guidotti2024counterfactual,lenatti2022novel}. \\
Fig.\ref{fig:trajectory} shows an example of the actual trajectory leading from a given factual to its related counterfactual. The trajectory can be retrieved by applying Algorithm 2. 
More in detail, for the factual in figure $x_\text{f}=(250,0,0)$ (thus we simulate a diabetic virtual subject, with no endogenous insulin production), by solving the optimization problem for $d=2$,  we get 
\begin{align}
v^*_{d=2}(\hat{x}) &=\zeta_1 + \zeta_2 \hat{x}_1 + \zeta_3 \hat{x}_2 + \zeta_4 \hat{x}_3 + \zeta_5 \hat{x}_1^2+ \zeta_6 \hat{x}_1\hat{x}_2\\ & +\zeta_7 \hat{x}_1 \hat{x}_3+ \zeta_8 \hat{x}_2^2 + \zeta_9 \hat{x}_2 \hat{x}_3 + \zeta_{10} \hat{x}_3^2
\end{align}

\begin{table}[!h]
\centering
\resizebox{0.2\textwidth}{!}{
\begin{tabular}{|c|c|}
\hline
\textbf{Coefficient} & \textbf{Value} \\
\hline
$\zeta_1$ & -2.7308 \\
$\zeta_2$ & 18.7159 \\
$\zeta_3$ & -353.7267 \\
$\zeta_4$ & 0.6326 \\
$\zeta_5$ & -27.9368 \\
$\zeta_6$ & 82.6978 \\
$\zeta_7$ & -2.0188 \\
$\zeta_8$ & -755356.4616 \\
$\zeta_9$ & -318.4366 \\
$\zeta_{10}$ & -0.48977 \\
$\gamma_1$ & -1.4879 \\
$\gamma_2$ & 4.7483 \\
$\gamma_3$ & 748.9628 \\
$\gamma_4$ & 2.3039 \\
\hline
\end{tabular}
}
\caption{Coefficients for $v^*_{d=2}(x)$ and $u^*_{d=2}(x)$}
\end{table}

From (\ref{u_star}), it is possible to derive
\begin{align}
u^*_{d=2}(\hat{x}) &= -\frac{1}{2} \nabla v^*(\hat{x})\cdot g(\hat{x}) \\   &= \gamma_1+ \gamma_2 \hat{x}_1+\gamma_3 \hat{x}_2+ \gamma_4 \hat{x}_3.
\end{align}
As described in Algorithm 2, the trajectory and the counterfactual $x_\text{cf}$ are retrieved by simulating in closed-loop the system  $\dot{\hat{x}} = \hat{f}(\hat{x}) + \hat{g}(\hat{x})u^*(\hat{x})$. \\

\newpage

\section{Uncertain systems}
This section is devoted to illustrate how the methodology previously discussed for fully parametrized systems can be generalized to the case of uncertain systems.
\subsection{Problem definition}
Consider a general time invariant  differential vector field of the form:
\begin{equation}
\dot{x}=f(x,u,\xi)
\label{vectorField}
\end{equation}
where $x \in \mathbb R^n$ is the state of the system, $u \in \mathbb{R}^m$ is the control and $\xi \in \mathbb{R}^{n_{\xi}}$ are the system parameters and \(f: \mathbb{R}^n \times \mathbb{R}^m \to \mathbb{R}^n\) is the vector field of the system dynamics.
We assume uncertainties on the knowledge of the parameters of the system as follows. \\
\textit{Assumption 6}. A subset $\zeta \subseteq \xi$ of the  parameters are unknown and take values in a set $\Sigma \subset \mathbb R^{n_{\zeta}}$. These values are considered to be time-independent and constant along trajectories.

For the system (\ref{vectorField}) we consider the general free terminal time optimal control problem of the form:
\begin{subequations}
\begin{align}
    J=&\inf_{u(\cdot),\ t^*}\Psi(x(t^*)) + \int_0^{t^*} \psi(x(t), u(t),\xi) \, dt \label{eq:objective_uncertain} \\
    \text{s.t.} \quad & \dot{x}(t) = f(x(t), u(t), \xi),  \label{eq:dynamics_uncertain} \\
    & x(0) \in X_0, \  x(t) \in X, \label{eq:state_constraints_uncertain} \\
    & x(t^*) \in X_T,\label{eq:terminal_condition_uncertain} \\
    & u(t) \in U, \label{eq:control_constraints_uncertain}\quad t \in [0, T],
\end{align}
\end{subequations}
where  \(t^*\) is the free terminal time, \(\Psi(x(t^*))\) is the terminal cost, \(\psi(x(t), u(t),\xi)\) is the stage cost.\\ $X_0,\ X, \ X_T$ constitute respectively the set of initial states, the set of the trajectories and the set of terminal states, $T$ is the time horizon. We will consider in general $X_0 \subset X$, $X_T \subseteq X$.\\
\textit{Remark}. Assumptions 1-5 previously made for the case of fully parametrized system continue to hold also for the case of uncertain systems.\\

To leverage the optimal control problem in the measures framework, following the developments in \cite{lasserre2008nonlinear}, \cite{henrion2012measures}, we will include the unknown parameters extending the state space representation of the system as follows:

\begin{equation}
\kappa:=[ x \quad \zeta_j]^{\text{T}}, \dot{\zeta_j}=0\\
\label{extended}
\end{equation}
for $j=1,\ldots, n_\zeta$. \\
In analogy with the classical definition of counterfactual and to what discussed above we will consider the \textit{factual} to be the initial condition of the general optimal control problem (\ref{eq:objective_uncertain})-(\ref{eq:control_constraints_uncertain}) applied to the system (\ref{extended}), $\kappa_\text{f} := \kappa(0)$. Similarly, we will indicate as \textit{counterfactual }the terminal state reached by the system in closed-loop with $u^*(\cdot) = \arg\min_{u(\cdot), t^*} J(\kappa(t),u(t))$, $\kappa_{\text{cf}}= \kappa(t^*\mid \kappa_\text{f},u^*)$. 
By means of the optimal control foundation, the concept of counterfactual is here endowed with the idea of \textit{minimum} effort for the system to access a safe target set.\\
We will denote with $\chi_0,\chi,\chi_T$ respectively the sets of initial conditions, of the state of trajectories and of terminal states of the extended system (\ref{extended}).

\textit{Remark}. By Assumption 2 , the sets $\chi_0,\chi,\chi_T$ inherit the algebraic properties of $X_0,X,X_T$.

\subsection{Problem solution in measures }
According to the developments in \cite{henrion2008nonlinear,henrion2012measures} the LP measure  program upper bounding the problem in  (\ref{eq:objective_uncertain})-(\ref{eq:control_constraints_uncertain}) reads as 
\begin{subequations}
\begin{align}
    p^* =\inf_{\mu_0,\mu,\mu_{t^*}} \quad & \langle \psi, \mu \rangle + \langle \Psi, \mu_{t^*} \rangle  \label{eq:functional_uncertain} \\
    \text{s.t.} \quad  & \delta_0 \otimes \mu_0 - \mu_{t^*} =\mathcal{L}^{*} \mu\label{eq:Liouville_uncertain} \\
    & \langle 1, \mu_0 \rangle = 1, \\
    & \mu \in \mathcal{M_+}([0,t^*] \times \chi \times U), \label{eq:support1_uncertain} \\
    &\mu_{t^*} \in  \mathcal{M_+}(\chi_T),\label{eq:support2_uncertain} 
\end{align}
\end{subequations}
where $\mu_0 = \delta_{\kappa=\kappa_\text{f}}$ is the initial measure  supported on the initial condition $\kappa_{\text{f}}$, $\mu$ is the occupation measure,$ \mu_{t^*} = \delta_{t^*} \otimes \delta_{\kappa=\kappa_{\text{cf}}}$ generalizes the concept of terminal measure with free terminal time. 
The infinite-dimensional optimization problem in the dual space of the continuous function is \cite{fantuzzi2020bounding,lasserre2008nonlinear}:
\begin{subequations}
\begin{align}
     q^* = \sup_{\gamma \in \mathbb{R}} \quad & \gamma
    \label{eq:functional_dual_uncertain} \\
    \text{s.t.} \quad  & \gamma \geq v(\kappa), \ \forall \kappa \in \chi_0 \\
    & \mathcal{L}_fv(\kappa) \leq -\psi(\kappa),\ \forall (\kappa,u) \in \chi \times U \notag\\
    & v(\kappa) \leq \Psi(\kappa),\ \forall \kappa \in \chi_T \label{eq:support_dual_uncertain}  \\
    & v \in C^1(\chi).\label{v_function_uncertain}
\end{align}
\label{eq:dual_uncertain}
\end{subequations}
The polynomial $v(\kappa(t))$  inner approximates the value function of system (\ref{extended}) along all the optimal trajectories of the system. \\
As design choice  for the development of our approach, we consider zero terminal cost ($\Psi =0$ ) and we consider the \textit{effort} (i.e., $\psi$) of the vector field to be the  ${L^2}$-norm of the control $u$.  Consequently, the problem ({\ref{eq:functional_uncertain})-(\ref{eq:support2_uncertain}}) results in

\begin{subequations}
\begin{align}
    p^* =\inf_{\mu_0,\mu,\mu_{t^*}} \quad & \langle u^2, \mu \rangle   \label{eq:functional_ac_uncertain} \\
    \text{s.t.} \quad  & \delta_0 \otimes \mu_0 - \mu_{t^*} = \mathcal{L}^{*}\mu \\
     & \langle 1, \mu_0 \rangle = 1, \label{deltaConstraint_uncertain}  \\
    & \mu \in \mathcal{M_+}([0,t^*] \times \chi \times U), \label{eq:support1_ac_uncertain} \\
    &\mu_{t^*} \in  \mathcal{M_+}(\chi_T) \label{eq:support2_Ac_uncertain} 
\end{align}
\label{eq:AC_uncertain}
\end{subequations}

Constraint (\ref{deltaConstraint_uncertain}) induces  an atomic representation for $\mu_0$. From (\ref{eq:Liouville_uncertain}) by choosing $v=1$ as test function it directly follows that also the final measure $\mu_{t^*}$ is atomic, being
\[
 \langle 1, \mu_{{0}} \rangle + \langle \mathcal{L}_f 1, \mu \rangle = \langle 1, \mu_{t^*} \rangle.
\]
More in detail, whenever (\ref{eq:AC_uncertain}) 
admits $s$ optimal solutions, then the final measure $\mu_{t^*}$ is s-atomic, i.e., it is a linear combination of atomic measures supported on s points. \\

\textit{Remark}. The case of completely known model (no uncertainty on model parameters) discussed above is a particular case of the more general one discussed here. In that case, the optimal control problem (and the related upper bounding program in measure space) refers to the original state space of the system. The uncertainties on parameters here results in a different optimization program in measures (\ref{eq:functional_ac_uncertain})-(\ref{eq:support2_Ac_uncertain}) with different decision variables and constraints.

\subsection{LMI formulation}  
The problem can be formulated in the truncated moment space (\ref{eq:functional_ac_uncertain})-(\ref{eq:support2_Ac_uncertain}) is solved as follows \cite{lasserre2008nonlinear,henrion2020moment}:

\begin{subequations}
\begin{align}
p^*_{d} = \ & \underset{y_0,y,y_{t^*}}{\text{min}} 
& & y_{02} \label{eq:cost_LMI_uncertain} \\
& \quad \text{s.t.}
& & 
{Liou_\alpha(y_0,y,y_{t^*})}\ \forall \alpha = 1, \ldots, 2d \label{eq:Liou_LMI_uncertain}\\ 
& & & y_{0,0}=1 \\
& & & \mathbb{M}_d(y_0),\ \mathbb{M}_d(y), \ \mathbb{M}_d(y_{t^*}) \succeq 0 \label{eq:2c_LMi_uncertain} \\
& & & \mathbb{M}_{d-d_k}(g_{0k}y) \succeq 0, \quad \forall k = 1 \ldots, N_0 \\
& & & \mathbb{M}_{d-d_k}(g_ky) \succeq 0, \quad \forall k = 1 \ldots, N \\
& & & \mathbb{M}_{d-d_k}(g_{t^*k}y_{t^*}) \succeq 0,   \quad \forall k = 1 \ldots, N_{t^*} \label{eq:2d_LMI_uncertain}
\end{align}
\label{eq:LMI_uncertain}
\end{subequations}

Measures $\mu_0, \mu, \mu_{t^*}$ have representing moment sequences respectively $y_0,y,y_{t^*}$ guaranteed by the algebraic properties of the support sets.
The cost is expressed as $2^{nd}$-order moment of the control $u$ with respect to the occupation measure $\mu$ ($y_{02})$.
${Liou_\alpha(y_0,y,y_{t^*})}$ denotes here Liouville's (\ref{eq:Liou_LMI_uncertain}) equation expressed in the moment space and ensures the relation 
\[
\langle \kappa^\alpha, \delta_0 \otimes \mu_0 \rangle +\langle \mathcal{L}_f \kappa^\alpha, \mu \rangle = \langle \kappa^\alpha, \mu_{t^*} \rangle.
\]
holds or all test functions $v(\kappa)=\kappa^\alpha$.
Moment and Localizing matrices in constraints (\ref{eq:2c_LMi_uncertain})-(\ref{eq:2d_LMI_uncertain}) represent the support set contraints (\ref{eq:support1_ac_uncertain})-(\ref{eq:support2_Ac_uncertain}) in the truncated moments space.

\textit{Remark}. In (\ref{vectorField}) we assume a time invariant vector field. 
The generality of the proposed approach is preserved for time varying systems solving the optimization problems by means of time dependent auxiliary function $v(t,\kappa(t))$ and extending the support measure constraints.
In this case, Liouville's equation in moment space results in 
\[
\langle \kappa^\alpha t^\beta, \delta_0 \otimes \mu_0 \rangle +\langle \mathcal{L}_f (\kappa^\alpha t^\beta, \mu \rangle = \langle \kappa^\alpha t^\beta, \mu_{t^*} \rangle
\]
for all test functions $v(\kappa,t)=\kappa^\alpha t^\beta$.
\subsection{Counterfactual Extraction}
\begin{algorithm}
\caption{Robust Counterfactual generation}
\begin{algorithmic}[1]
\Require Sets \(\chi_0\), \(\chi\), \(\chi_T\), vector field \(f\), cost functional,  degree \(d\)
\Ensure Robust Counterfactual extraction \\ 
Define the known support components of  $\mu_0$, spt$(\mu_0)=\kappa_\text{f}$
  \State Solve (\ref{eq:AC_uncertain}) at degree \(d\)
  \State Extract $y_{1,t^*}$ from $\mathbb{M}_d(y_{t^*})$
  \State $\kappa_\text{cf}=y_{1,t^*}$   
\end{algorithmic}
\end{algorithm}

This section shows {Algorithm 1} previously proposed can be extended ti the generalized robust framework to allow the extraction of counterfactuals. 
We extend {Algorithm 1} as follows: \\
\textit{Proposition 3}. Consider the problem 
in (\ref{eq:functional_ac_uncertain})-(\ref{eq:support2_Ac_uncertain}). Assume it is  feasible and its solution exists for all suitable orders $d$. Then, the counterfactual $\kappa_\text{cf}=\text{spt}(\mu_{t^*}=\delta_{\kappa=\kappa_{t^*}})$ associated with the factual $\kappa_\text{f}=\text{spt}(\mu_0=\delta_{\kappa=\kappa_\text{f}})$ can be retrieved from the moment matrix and specifically it holds that
$\kappa_\text{cf}=y_{1,t^*}$, where $y_{1,t^*}$ is the $1^{st}$-order moment of the final measure $\mu_{t^*}$. 

Algorithm 3 summarizes the procedure. 
As already mentioned, there are no theoretical guarantees on the convergence of the solution for finite relaxation order $d$ in presence of constraints related to the dynamics (i.e., Liouville's equation). More in detail, finite relaxation order solutions provide lower bounds  $p^*_{d}=q^*_{d} \leq p^*=q^*$ and the convergence is only guaranteed as $d \to \infty$. \\
The formulation we design in this study inherently yields a safety problem. The formal results expressed by \textit{Theorem 1},\textit{Theorem 2}, \textit{Theorem 3} hold also for the case here studied of uncertainties on the model for the extended system (\ref{extended}).

\textit{Remark}. Solving the dual program (\ref{eq:functional_dual_uncertain})-(\ref{v_function_uncertain}), counterfactuals can be equivalently extracted retrieving  $u^*(\kappa) := \arg\min_{u \in U} \left\{ \mathcal{L}_f v(\kappa) 
+ \psi(\kappa, u) \right\}$ and simulating the closed-loop system via Algorithm 2.
\subsection{RESULTS}
\begin{figure}[!h]
\centering
{\includegraphics[scale=0.5]{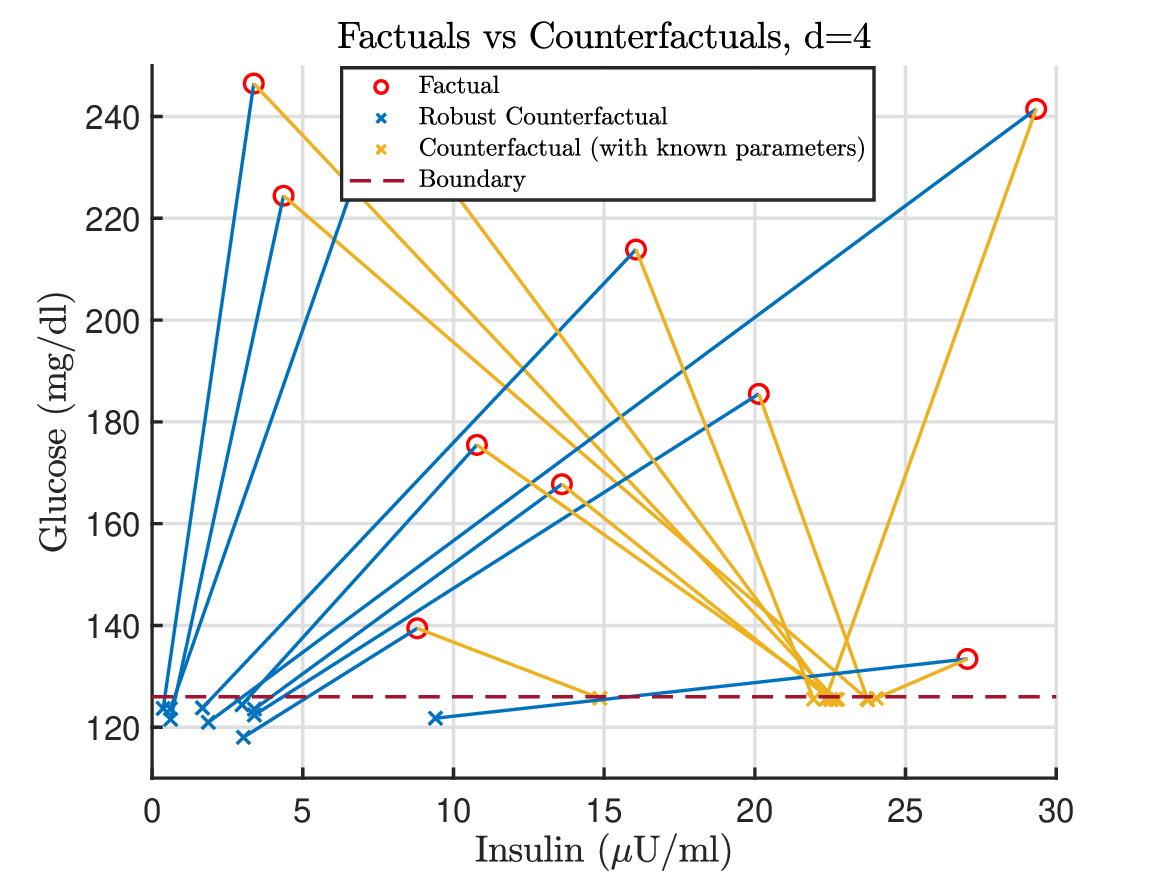}}
{\includegraphics[scale=0.5]{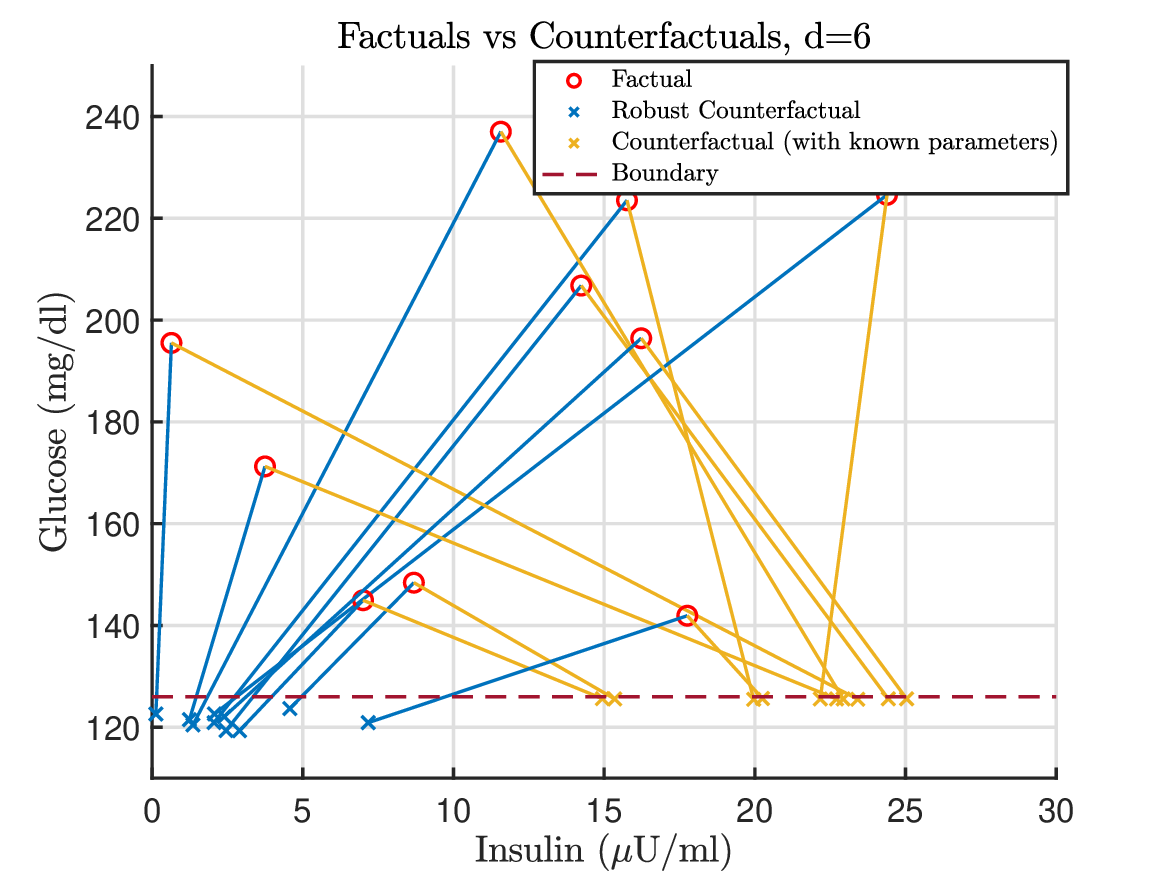}}
\caption{Factual-counterfactual pairs as obtained by solving the optimization problem in the moment space with relaxation order $d=4$ (top panel) and $d=6$ (bottom panel). Robust counterfactuals lie at lower values of $G$ with increased average distance from the boundary.}
\label{fig:d4}
\end{figure}

\begin{figure}[!h]
\centering
{\includegraphics[scale=0.5]{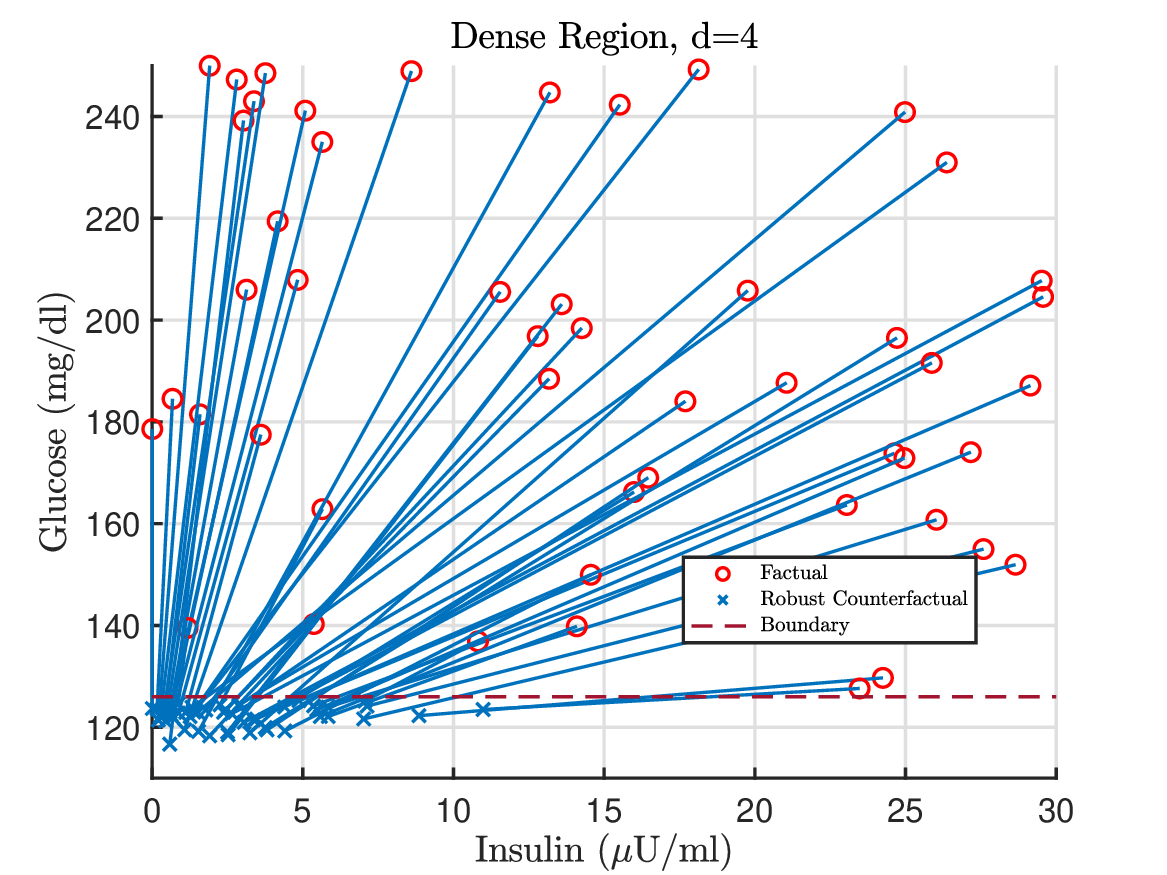}}
{\includegraphics[scale=0.5]{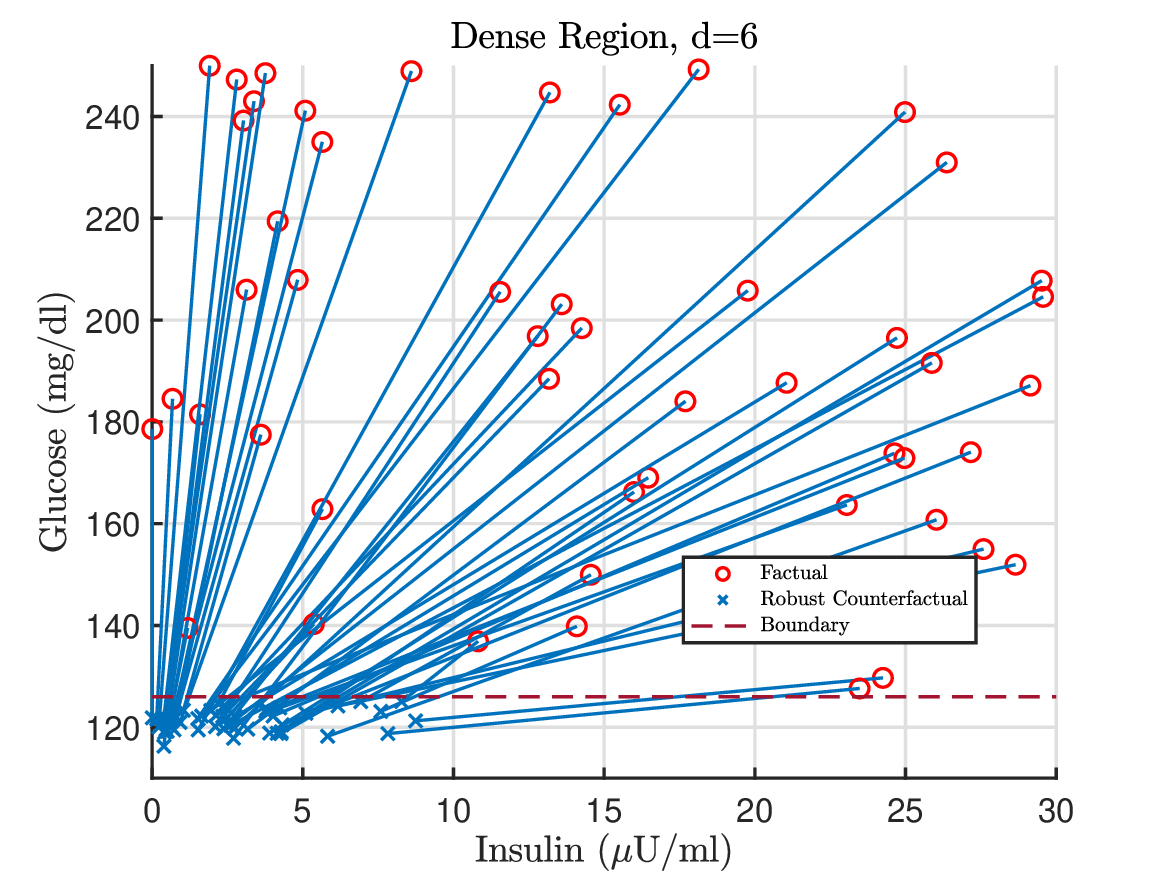}}
\caption{Factuals vs Counterfactuals obtained for $d=4$ (top panel) and $d=6$ (bottom panel). Both the plots highlight the same dense region of counterfactuals within the interval of values of $I$ between $0$ and $10$}
\label{fig:DenseRegions}
\end{figure}

We apply the proposed approach to the system (\ref{Bergman}) scaled as (\ref{Bergman_scaled}), where 
we assume that parameters ($p_2$,$p_3$) are unknown, $p_2\in [0.049,0.051]$ and $p_3 \in [2.7 \cdot 10^{-5},3 \cdot 10^{-5}]$.
Indeed this is a reasonable choice, as parameters $p_3$ and $p_3$ are related to the dynamics of the remote insulin ($x_2$) hence identifying their values is not trivial in a real-world scenario.
We assume the safe set to be  $X_T = \{{x} = (x_1, x_2, x_3) \in \mathbb{R}^3 \mid 80 \leq x_1 < 126 \}$ whereas  initial conditions are sampled within the set $X_0 = \{{x} = (x_1, x_2, x_3) \in \mathbb{R}^3  \mid (126 \leq x_1 \leq 260), (0\leq x_3 \leq 30)\} $\cite{BergmanEtAl1979}. \\
In the following we will compare the results obtained when the system is assumed to be known, as discussed above, and the case when the system is assumed to be uncertain.
Robust Counterfactuals for the uncertain system are extracted via Algorithm 3 leveraging the proposed methodology.
Fig.\ref{fig:d4} shows pairs of factuals and counterfactuals when $10$ factuals are randomly sampled over the unsafe set and the optimization problem is solved for $d=4$ (top panel) and $d=6$ (bottom panel) both for the known system and the uncertain one. \\
For a given factual (initial, unsafe condition for the system, red dot), the plot highlights in yellow and blue respectively the association between the counterfactual extracted in case of known system and  the counterfactual robust with respect to the model uncertainties.  
As it can be observed from the plot, when the system is assumed known, counterfactuals tend to lie on the boundary (dashed line) separating the two sets. Conversely, in the robust scenario, counterfactuals lie in the safe set in points at lower values of $G$.
This result appears to be reasonable, as the robust optimal control law ensures that the system keeps safe despite the uncertainties on its dynamics and this results in an increased average distance from the boundary.\\
It should be remarked that the generated counterfactuals are points reached by the system dynamics with lower bound with respect to the cost achieved as $d \to \infty$.
Indeed Fig. \ref{fig:DenseRegions} shows the results obtained when robust counterfactuals are extracted solving the optimization problem for $d=4$ (top panel) and $d=6$ (bottom panel) and factuals are $50$ randomized points over the space of the initial conditions.
As it can be observed, the plot highlights a specific property of counterfactuals, as they tend to cluster in dense regions of the state space, independently of the relaxation order $d$ that solves that optimization problem as observed also previously for the case of fully parametrized system.
It can be easily observed that in Fig. \ref{fig:DenseRegions} the dense region lies withing an interval for the values of $I$ between 0 and 10 $\mu$U/ml and this dense region is different with respect to that denoted in Fig. \ref{fig:DenseRegions_known} when the system is assumed to be known.
It would be of interest to investigate to what extent the presence of dense regions is in general associated with specific properties of the system under consideration.

\section{CONCLUSION}
This work represents a novel proposal aimed at introducing the concept of counterfactuals in the field of control and its applications, leveraging the powerful tools of occupation measures and moment-SOS hierarchy.
Control-driven counterfactuals are inherently informed about the system dynamics, thus embedding the information about how the system can be steered to the safe set and which terminal states can be reached at a given terminal time. This distinctive feature represents a conceptual step forward with respect to the AI-framework, where most counterfactuals overlook the knowledge of the system dynamics and may represent points not reachable by the system in a real-world scenario.
Further developments will investigate whether some of the formal properties described in  \cite{guidotti2024counterfactual} and \cite{carlevaro2023multi} can be extended to the case of control-driven robust counterfactuals. Moreover, future work will deal with a stochastic oriented setting, leveraging different initial distributions and final distributions of counterfactuals.
Within the context of life science applications, this study set the basis for future works leveraging the proposed approach of control-driven counterfactuals to integrate control and AI with the aim of feeding machine learning algorithms with the physics knowledge acquired via the system framework, with specific reference to applications in the field of type 2 diabetes control \cite{lenatti2022novel,DePaolaEtAl,de2024novel}.
\section{ACKNOWLEDGMENTS}
The authors sincerely thank Mario Sznaier and Costantino Lagoa for the very fruitful discussion about occupation measures. \\
This work was supported in part by the European Union through the Project PRAESIIDIUM “Physics Informed Machine Learning-Based Prediction and Reversion of Impaired Fasting Glucose Management" (call HORIZON-HLTH-2022-STAYHLTH-02), Grant 101095672. Views and opinions expressed are however those of the authors only and do not necessarily reflect those of the European Union or the European Health and Digital Executive Agency (HADEA). Neither the European Union nor the HADEA can be held responsible for them.\\ This work was carried out within the Italian National Ph.D. Program in Autonomous Systems (DAuSy), coordinated by Polytechnic of Bari, Italy
\\ 

\bibliography{Counterfactuals_arxiv}

\begin{thebibliography}{10}

\bibitem{ecc2025}
P.~De~Paola, J.~Miller, A.~Borri, A.~Paglialonga, and F.~Dabbene, ``A control system framework for counterfactuals: an optimization-based approach,'' in {\em Submitted to Proceedings of the European Control Conference 2025 (ECC)}, 2025.

\bibitem{Rocond2025cf}
P.~De~Paola, J.~Miller, A.~Borri, A.~Paglialonga, and F.~Dabbene, ``Robust control-driven counterfactual generation for uncertain systems,'' in {\em Submitted to Proceedings of the 11th IFAC Symposium on Robust Control Design (ROCOND’25)}, 2025.

\bibitem{guidotti2024counterfactual}
R.~Guidotti, ``Counterfactual explanations and how to find them: literature review and benchmarking,'' {\em Data Mining and Knowledge Discovery}, vol.~38, no.~5, pp.~2770--2824, 2024.

\bibitem{lewis2013counterfactuals}
D.~Lewis, {\em Counterfactuals}.
\newblock John Wiley \& Sons, 2013.

\bibitem{wachter2017counterfactual}
S.~Wachter, B.~Mittelstadt, and C.~Russell, ``Counterfactual explanations without opening the black box: Automated decisions and the gdpr,'' {\em Harv. JL \& Tech.}, vol.~31, p.~841, 2017.

\bibitem{carlevaro2022counterfactual}
A.~Carlevaro, M.~Lenatti, A.~Paglialonga, and M.~Mongelli, ``Counterfactual building and evaluation via explainable support vector data description,'' {\em IEEE Access}, vol.~10, pp.~60849--60861, 2022.

\bibitem{lenatti2022novel}
M.~Lenatti, A.~Carlevaro, A.~Guergachi, K.~Keshavjee, M.~Mongelli, and A.~Paglialonga, ``A novel method to derive personalized minimum viable recommendations for type 2 diabetes prevention based on counterfactual explanations,'' {\em Plos one}, vol.~17, no.~11, p.~e0272825, 2022.

\bibitem{carlevaro2023multi}
A.~Carlevaro, M.~Lenatti, A.~Paglialonga, and M.~Mongelli, ``Multi-class counterfactual explanations using support vector data description,'' {\em IEEE Transactions on Artificial Intelligence}, 2023.

\bibitem{stepin2021survey}
I.~Stepin, J.~M. Alonso, A.~Catala, and M.~Pereira-Fari{\~n}a, ``A survey of contrastive and counterfactual explanation generation methods for explainable artificial intelligence,'' {\em IEEE Access}, vol.~9, pp.~11974--12001, 2021.

\bibitem{kment2006counterfactuals}
B.~Kment, ``Counterfactuals and explanation,'' {\em Mind}, vol.~115, no.~458, pp.~261--310, 2006.

\bibitem{henrion2008nonlinear}
D.~Henrion, J.~B. Lasserre, and C.~Savorgnan, ``Nonlinear optimal control synthesis via occupation measures,'' in {\em 2008 47th IEEE Conference on Decision and Control}, pp.~4749--4754, IEEE, 2008.

\bibitem{lasserre2008nonlinear}
J.~B. Lasserre, D.~Henrion, C.~Prieur, and E.~Tr{\'e}lat, ``Nonlinear optimal control via occupation measures and lmi-relaxations,'' {\em SIAM journal on control and optimization}, vol.~47, no.~4, pp.~1643--1666, 2008.

\bibitem{henrion2020moment}
D.~Henrion, M.~Korda, and J.~B. Lasserre, {\em Moment-sos Hierarchy, The: Lectures In Probability, Statistics, Computational Geometry, Control And Nonlinear Pdes}, vol.~4.
\newblock World Scientific, 2020.

\bibitem{lasserre2009moments}
J.~B. Lasserre, {\em Moments, positive polynomials and their applications}, vol.~1.
\newblock World Scientific, 2009.

\bibitem{MillerEtAl}
J.~Miller, D.~Henrion, and M.~Sznaier, ``Peak estimation recovery and safety analysis,'' {\em IEEE Control Systems Letters}, vol.~5, no.~6, pp.~1982--1987, 2021.

\bibitem{case2024}
P.~De~Paola, A.~Borri, A.~Paglialonga, P.~Palumbo, and F.~Dabbene, ``Polynomial approximation of regions of attraction via occupation measures: an application to a biological autonomous system,'' in {\em Proceedings of the IEEE 20th International Conference on Automation Science and Engineering (CASE)}, 2024.

\bibitem{nash1987linear}
P.~Nash and E.~J. Anderson, ``Linear programming in infinite-dimensional spaces: theory and applications,'' {\em (No Title)}, 1987.

\bibitem{fattorini1999infinite}
H.~O. Fattorini, {\em Infinite dimensional optimization and control theory}, vol.~54.
\newblock Cambridge University Press, 1999.

\bibitem{khalil2002nonlinear}
H.~Khalil, {\em Nonlinear systems}.
\newblock Prentice Hall, 2002.

\bibitem{BergmanEtAl1979}
R.~N. Bergman, Y.~Z. Ider, C.~R. Bowden, and C.~Cobelli, ``Quantitative estimation of insulin sensitivity.,'' {\em American Journal of Physiology-Endocrinology And Metabolism}, vol.~236, no.~6, p.~E667, 1979.

\bibitem{WhelanEtAl2022}
M.~Whelan and L.~Bell, ``The english national health service diabetes prevention programme (nhs dpp): A scoping review of existing evidence,'' {\em Diabetic Medicine}, vol.~39, no.~7, p.~e14855, 2022.

\bibitem{henrion2012measures}
D.~Henrion, M.~Ganet-Schoeller, and S.~Bennani, ``Measures and lmi for space launcher robust control validation,'' {\em IFAC Proceedings Volumes}, vol.~45, no.~13, pp.~236--241, 2012.

\bibitem{fantuzzi2020bounding}
G.~Fantuzzi and D.~Goluskin, ``Bounding extreme events in nonlinear dynamics using convex optimization,'' {\em SIAM journal on applied dynamical systems}, vol.~19, no.~3, pp.~1823--1864, 2020.

\bibitem{DePaolaEtAl}
P.~F. De~Paola, A.~Paglialonga, P.~Palumbo, K.~Keshavjee, F.~Dabbene, and A.~Borri, ``The long-term effects of physical activity on blood glucose regulation: a model to unravel diabetes progression,'' {\em IEEE Control Systems Letters}, vol.~7, pp.~2916--2921, 2023.

\bibitem{de2024novel}
P.~F. De~Paola, A.~Borri, F.~Dabbene, K.~Keshavjee, P.~Palumbo, and A.~Paglialonga, ``A novel mathematical model for predicting the benefits of physical activity on type 2 diabetes progression,'' {\em arXiv preprint arXiv:2404.14915}, 2024.

\end{thebibliography}
\bibliographystyle{ieeetr}
\end{document}